\documentclass[12pt]{article}
\usepackage[top = 2 cm, bottom = 3 cm, left = 3 cm, right = 2 cm]{geometry}
\usepackage{authblk, cite, color, amssymb, amsmath}
\usepackage[colorlinks = true, linkcolor = blue, citecolor = red]{hyperref}
\usepackage[dvipsnames]{xcolor}

\begin{document}

\title{\bf  Dirac and Maxwell equations in Split Octonions}

\author{R. Beradze and T. Shengelia}
\affil{\small Javakhishvili Tbilisi State University, 3 Chavchavadze Avenue, Tbilisi 0179, Georgia\\
E-mail: {\it revazberadze@gmail.com} and {\it cotneshengelia@yahoo.com}}

\maketitle

\begin{abstract}
The split octonionic form of Dirac and Maxwell equations are found. In contrast with the previous attempts these equations are derived from the octonionic analyticity condition and also we use different basis of the 8-dimensional space of split octonions.

\vskip 1mm

\end{abstract}
\vskip 5mm


\section{Introduction}

Octonions, as they form the widest normed division algebras after real numbers, complex numbers and quaternions \cite{Sc, Sp-Ve, Baez}, are interesting mathematical objects for physical applications \cite{Rev}. In this paper our purpose is to rewrite Dirac and Maxwell equations in vacuum using split octonions. Octonions are 8-dimensional number and that's why instead of system of equations, Dirac, as well as Maxwell equation, each will be represented with only one equation. This will simplify applying mathematical operations and may lead to new physical results.

\section{Octonionic geometry}

Algebra of the eight basis elements of split-octonions, we use in this paper, is similar as it was derived in \cite{Gog, geo}:
\begin{eqnarray} \label{algebra}
&J_n^2 = 1~, ~~~~~ j_n^2=-1~, ~~~~~ I^2=1~,\nonumber\\
&j_n = \frac{1}{2} \varepsilon_{nmk}J^mJ^k~, ~~~~~ I = J_1j_1 = J_2j_2 = J_3j_3 ~,\nonumber \\
&J_nJ_m = \varepsilon_{nmk} j^k = - J_mJ_n ~, ~~~ j_nj_m = \varepsilon_{nmk} j^k = -j_mj_n ~, ~~~ j_mJ_n  = \varepsilon_{nmk}J^k = - J_nj_m ~, \\
&J_nI = j_n = - IJ_n ~, ~~~~~ j_nI = J_n = -Ij_n ~,\nonumber\\
&J_n^\dag = - J_n ~, ~~~~~ j_n^\dag  = \frac 12 \varepsilon_{nmk}J^{k\dag}J^{m\dag} = - j_n~, ~~~~~ I^\dag = J_3^\dag J_2^\dag J_1^\dag= - I~. \nonumber
\end{eqnarray}
where n,m,k = 1,2,3 and $\varepsilon_{nmk}$ is the totally antisymmetric unit tensor.

We assume that world-lines of particles can be parameterized using the split octonion elements \cite{Gog, geo},
\begin{equation} \label{s}
s = \omega + \lambda^nJ_n + x^nj_n + t I~, ~~~~~ (n = 1, 2, 3)
\end{equation}
where $x^nj_n = \delta_{nm}x^nj^m$ and $\delta^{nm}$ is Kronecker's delta. Here $t$ and $x^n$ denote ordinary space-time coordinates and $\omega$ and $\lambda^n$ are interpreted as classical action and wavelength associated with octonionic signals \cite{Gog, geo}. Using the algebra (\ref{algebra}) conjugation of (\ref{s}) will give:
\begin{equation} \label{s*}
s^\dag = \omega - \lambda_n J^n - x_n j^n - t I~.
\end{equation}
Thus the norm of (\ref{s}), space-time interval, has the form:
\begin{equation} \label{sN}
N^2 = ss^\dag = s^\dag s = \omega^2 - \lambda^2 + x^2 - t^2 ~.
\end{equation}
It represents the 8-dimensional space-time with (4+4) signature and reduces to ordinary Minkowski space if $\omega^2 - \lambda^2 = 0$. As for classical case we demand (\ref{sN}) to be non-negative and also for physical events vector-like part of (\ref{s}) to be time-like, i.e. $t^2 + \lambda_n\lambda^n > x_nx^n~$ \cite{geo}.

The norm (\ref{sN}) can be viewed as some kind of space-time interval with three extra time-like dimensions. The ordinary time parameter, $t$, corresponds to the distinguished octonionic basis unit, $I$, while the other three time-like parameters, $\lambda_n$, have a natural interpretation as wavelengths, i.e. do not relate to time as conventionally understood. Within this picture, in front of time-like coordinates in the expression of pseudo-Euclidean octonionic intervals there naturally appear two fundamental physical parameters, the light speed and Planck's constant. Then from the requirement of positive definiteness of norms, together with the introduction of the maximal velocity, there follow conditions which are equivalent to uncertainty relations \cite{Gog, geo}. Also it is known that a unique physical system in multi-time formalism generates a large variety of 'shadows' (different dynamical systems) in (3+1)-subspace \cite{times}. One can speculate that  information of multi-dimensional structures, which is retained by these images of the initial system, might takes the form of hidden symmetries in the octonionic particle Lagrangians \cite{geo}.

Split algebras contain special elements with zero norms (zero divisors) \cite{Sc}, which are important structures in physical applications \cite{So}. For the coordinate function (\ref{s}) we can define the deferential zero divisor,
\begin{equation} \label{d}
\frac {d}{ds} = \frac 12 \left[ \frac {d}{d\omega} + J_n \frac {d}{d\lambda_n} + j_n \frac {d}{dx_n} + I\frac {d}{dt}\right]~,
\end{equation}
such that its action upon $s$ is:
\begin{equation}
\frac {ds}{ds} = 1~.
\end{equation}
The operator (\ref{d}) annihilates $s^\dag$, while the conjugated derivative operator,
\begin{equation} \label{d*}
\frac {d}{ds^\dag} = \frac 12 \left[ \frac {d}{d\omega} - J_n \frac {d}{d\lambda_n} - j_n \frac {d}{dx_n} - I\frac {d}{dt}\right]~,
\end{equation}
is zero divisor for $s$, i.e.
\begin{equation}
\frac {ds^\dag}{ds} = \frac {ds}{ds^\dag} = 0~.
\end{equation}
From these relations it is clear that the interval (\ref{sN}) is a constant function for the restricted left octonionic gradient operators,
\begin{eqnarray}
\frac {d}{ds_L} \left(s^\dag s\right) = \left(\frac {ds^\dag}{ds} \right) s = 0~, \nonumber \\
\frac {d}{ds^\dag_L} \left(ss^\dag\right) = \left(\frac {ds}{ds^\dag} \right) s^\dag = 0~,
\end{eqnarray}
and the invariance of the intervals,
\begin{equation}
ds^2 = ds ds^\dag = ds^\dag ds~,
\end{equation}
in the space of split octonions can be viewed as an algebraic property.

The octonionic particle wavefunctions $\Psi$, in general, should depend on $s$ and on $s^\dag$ as well. Thus we need the condition of analyticity of functions of split octonionic variables,
\begin{equation} \label{dPsi/ds}
\frac {d\Psi(s,s^\dag)}{ds^\dag} = 0~,
\end{equation}
which is similar to the Cauchy-Riemann equations from complex analysis.

Now let us show that the system of eight real-valued algebraic conditions (\ref{dPsi/ds}), in certain cases \cite{Gog-Split}, lead to the octonionic Maxwell and Dirac equations \cite{Dir, Max}.


\section{The Dirac equation}

Let us consider the Dirac equation,
\begin{equation} \label{dirac}
\left(i\gamma^\nu \partial_\nu  - m \right) \Psi = 0~,
\end{equation}
using the standard representation for the gamma-matrices,
\begin{equation}
\gamma^0 =
\begin{pmatrix}
{\bf 1} & 0\cr
0 & -{\bf 1}
\end{pmatrix}~, ~~~~~
\gamma^i =
\begin{pmatrix}
0 & \sigma^i\cr
-\sigma^i & 0
\end{pmatrix}~, ~~~~~
\end{equation}
where $\sigma^i$ are Pauli matrices and $\bf 1$ denotes the unit $2 \times 2$ matrix. The 4-dimensional complex spinor in (\ref{dirac}) can be written in the form \cite{Dir}:
\begin{equation} \label{Psi}
\Psi = \left(
\begin{array}{c}
-y_0 + i l_3 \\
l_2 - i l_1 \\
y_3 + i l_0 \\
y_1 + i y_2
\end{array} \right) ~,
\end{equation}
where eight scalar functions $y^\nu$ and $l^\nu$ ($\nu = 0, 1, 2, 3$) are written without any explanation of their nature. Then introducing the notations,
\begin{eqnarray} \label{LY}
L_{\nu\mu} &=& \partial_\nu l_\mu - \partial_\mu l_\nu ~, ~~~~~(\nu, \mu = 0,1,2,3) \nonumber\\
Y_{\nu\mu} &=& \partial_\nu y_\mu - \partial_\mu y_\nu ~,
\end{eqnarray}
from (\ref{dirac}) and (\ref{Psi}) we can get the set of eight real-valued differential equations:
\begin{eqnarray} \label{system}
L_{30} + Y_{12} &=& -m l_3 ~, ~~~~~  -\partial^\nu y_\nu = - m y_0 ~,  \nonumber\\
L_{02} + Y_{13} &=& m l_2 ~, ~~~~~ L_{10} + Y_{23} = - m l_1 ~, \nonumber \\
L_{21} + Y_{30} &=& - m y_3 ~, ~~~~~ -\partial^\nu l_\nu = - m l_0 ~,  \\
L_{32} + Y_{10} &=& - m y_1~, ~~~~~ L_{13} + Y_{20} = - m y_2 ~, \nonumber
\end{eqnarray}
where we use the Minkowski metric with the signature $(+---)$ for summing.

Now we want to show that the system (\ref{system}) is equivalent to the octonionic Cauchy-Riemann condition (\ref{dPsi/ds}). As in (\ref{Psi}) we use the eight real functions $y^\nu$ and $l^\nu$ to define the octonionic wave function,
\begin{equation} \label{psi}
\Psi =  l_0 + y_ij^i + l_iJ^i + y_0 I ~. ~~~~~(i = 1,2,3)
\end{equation}
For simplicity we suppose that $y^\nu$ and $l^\nu$ are independent of the three octonionic coordinates $\lambda^i$. Then the condition of octonion analyticity (\ref{dPsi/ds}) takes the form:
\begin{equation} \label{continuity}
\frac {d\Psi}{ds^\dag} = \left( \partial_\omega - \nabla \right) \Psi = 0 ~,
\end{equation}
where we introduced the analog of 4-dimensional d'Alembertian,
\begin{equation} \label{nabla}
\nabla = \left(\partial_t + J^n\partial_n \right)I ~.
\end{equation}
We also suppose that $y^\nu$ and $l^\nu$ depend on $\omega$ only via the factor $e^{m\omega}$. Then using the algebra of the octonionic basis elements (\ref{algebra}), and the notation (\ref{LY}), we get:
\begin{eqnarray} \label{nablaPsi}
e^{m\omega}\left[\left(\partial^\nu l_\nu - m l_0\right)  + \left( L^{i0} + \epsilon^{ijk} Y_{jk} - m l^i\right)J_i ~+ \right. \nonumber \\
\left. + \left( Y^{i0} - \epsilon^{ijk} L_{jk} - m y^i \right) j_i + \left( \partial^\nu y_\nu - m y_0\right) I \right]= 0 ~,
\end{eqnarray}
where $\nu = 0,1,2,3$ and $i, j, k = 1, 2, 3$ .
Equating to zero coefficients in front of octonionic basis units in (\ref{nablaPsi}) we have a system of eight equations, which resembles to (\ref{system}).

\section{ The Maxwell equations}

Similar to \cite{Max}, but using different basis (\ref{s}), we define the octonionic electromagnetic potential as:
\begin{equation} \label{A}
A = b + J_n B^n + j_n A^n + I \varphi ~, ~~~~~~(n = 1, 2, 3)
\end{equation}
where $\varphi$ and $A^n$ are the components of the ordinary Maxwell's electromagnetic 4-vector. Appearance of the terms $b$ and $B^n$ in (\ref{A}) is related with existence of the extra degrees of freedom in the octonionic algebra, but here we don't want to specify their physical meaning, we just want to derive the Maxwell classical equations in some approximation.

As our aim is to derive classical equations, we can ignore derivatives of the octonionic vector-potential (\ref{A}) by the four coordinates, $\omega$ and $\lambda^n$, of the split octonion 8-space (\ref{s}) and use the 4-dimensional d'Alembertian (\ref{nabla}) to write the electro-magnetic field:
\begin{equation} \label{F}
F = \nabla A = J_n (E^n - \widehat{H}^n) + j_n (\widehat{E}^n - H^n)~,
\end{equation}
where,
\begin{eqnarray}
\widehat{E}^n = \partial_n b - \partial_t B^n ~, ~~~~~ \widehat{H}^n = \epsilon^{n m k} \partial_m B_k~, \nonumber \\
E^n = \partial_n \varphi - \partial_t A^n~, ~~~~~ H^n = \epsilon^{n m k} \partial_m A_k ~.
\end{eqnarray}
Also postulated the Lorenz gauge,
\begin{equation}
\partial_t b - \partial_n B^n = \partial_t \varphi - \partial_n A^n = 0~.
\end{equation}
Then the conditions of analyticity (\ref{dPsi/ds}) gives:
\begin{eqnarray} \label{octonionic Maxwell Full}
\nabla F = \partial_n\left ( -\widehat{E}^n + H^n\right ) + J_n \left [\partial_t \left ( -\widehat{E}^n + H^n \right ) +
\epsilon^{n m k} \partial_m \left ( -E_k + \widehat{H}_k \right ) \right ] + \nonumber \\
+ j_n \left [\partial_t \left ( -E^n + \widehat{H}^n\right ) +
\epsilon^{n m k} \partial_m \left (\widehat{E}_k - H_k \right ) \right ]+ I \partial_n \left ( -E^n + \widehat{H}^n\right ) = 0 ~.
\end{eqnarray}
If we assume that $b$ and $B$ are constants, or small
\begin{equation}
|b|, |B^n| \ll |\varphi|, |A^n|~,
\end{equation}
and neglect $\widehat{E}^n$ and $\widehat{H}^n$, we get
\begin{equation} \label{octonionic Maxwell}
\partial_n H^n + J_n \left ( \partial_t H^n - \epsilon^{n m k} \partial_m E_k \right ) - j_n \left (\partial_t E^n + \epsilon^{n m k} \partial_m H_k \right ) - I \partial_n E^n = 0 ~.
\end{equation}
Equating to zero coefficients in front of the octonionic basis units in (\ref{octonionic Maxwell}) results the full set of the ordinary homogeneous Maxwell equations.


\section{Conclusion}

In this paper the split octonionic form of Dirac and Maxwell equations are found. Unlike to previous similar attempts we derived these equations from the octonionic analyticity condition, which is similar to the Cauchy-Riemann equations in complex analysis. Another novelty is that we use different basis of the 8-dimensional space of split octonions.



\begin{thebibliography}{9}

\bibitem{Sc} R. Schafer,
             {\it Introduction to Non-Associative Algebras}
             (Dover, NY 1995).

\bibitem{Sp-Ve} T.A. Springer and F.D. Veldkamp,
               {\it Octonions, Jordan Algebras and Exceptional Groups}, Springer Monographs in Mathematics
               (Springer, Berlin 2000).

\bibitem{Baez} J.C. Baez,
              Bull. Am. Math. Soc. {\bf 39} (2002) 145,	arXiv: math/0105155 [math.RA].

\bibitem{Rev} S. Okubo,
             {\it Introduction to Octonion and Other Non-Associative Algebras in Physics}
             (Cambrodge Univ. Press, Cambridge 1995); \\
              D. Finkelstein,
             {\it Quantum Relativity: A Synthesis of the Ideas of Einstein and Heisenberg}
             (Springer, Berlin 1996); \\
              F. G\"{u}rsey and C. Tze,
             {\it On the Role of Division, Jordan and Related Algebras in Particle Physics}
             (World Scientific, Singapore 1996); \\
              J. L\~{o}hmus, P. Paal and L. Sorgsepp,
             {\it Nonassociative Algebras in Physics}
             (Hadronic Press, Palm Harbor 1994);
             Acta Appl. Math. {\bf 50} (1998) 3.

\bibitem{Gog} M. Gogberashvili,
             arXiv: hep-th/0212251;
             Adv. Appl. Clif. Alg. {\bf 15} (2005) 55, arXiv: hep-th/0409173;
             Adv. Math. Phys. {\bf 2009} (2009) 483079, arXiv: 0808.2496 [math-ph];
             Progr. Phys. {\bf 12} (2016) 30, arXiv: 1511.05818 [physics.gen-ph];
             Int. J. Geom. Meth. Mod. Phys. {\bf 13} (2016) 1650092, arXiv: 1602.07979 [physics.gen-ph].

\bibitem{geo} M. Gogberashvili and  O. Sakhelashvili,
              Adv. Math. Phys. {\bf 2015} (2015) 196708, arXiv: 1506.01012 [math-ph].

\bibitem{times} E.A.B. Cole,
               Nuov. Cim. {\bf A 40} (1977) 171;
               J. Phys. {\bf A 13} (1980) 109; \\
                M. Gogberashvili,
              Phys. Lett. {\bf B 484} (2000) 124, arXiv: hep-ph/0001109; \\
               M. Gogberashvili and P. Midodashvili,
              Phys. Lett. {\bf B 515} (2001) 447, arXiv: hep-ph/0005298;
              Europhys. Lett. {\bf 61} (2003) 308, arXiv: hep-th/0111132; \\
               J. Christian,
              Int. J. Mod. Phys. {\bf D 13} (2004) 1037, arXive: gr-qc/0308028;
              in {\it Relativity and the Dimensionality of the World}, ed. V. Petkov
              (Springer, NY 2007), arXiv: gr-qc/0610049; \\
               M.V. Velev,
              Phys. Essays {\bf 25} (2012) 3.

\bibitem{So} A. Sommerfeld,
             {\it Atombau und Spektrallinien}, II Band
             (Vieweg, Braunschweig 1953).

\bibitem{Gog-Split} M. Gogberashvili,
                   Eur. Phys. J. {\bf C 74} (2014) 3200, arXiv: 1410.4136 [physics.gen-ph].

\bibitem{Dir} M. Gogberashvili,
              Int. J. Mod. Phys. {\bf A 21} (2006) 3513, arXiv: hep-th/0505101.

\bibitem{Max} M. Gogberashvili,
              J. Phys. {\bf A 39} (2006) 7099, arXiv: hep-th/0512258.

\end{thebibliography}
\end{document}